\pdfoutput=1

\documentclass[11pt]{article}

\usepackage[final]{acl}

\usepackage{times}
\usepackage{latexsym}
\usepackage{amsmath}
\usepackage[T1]{fontenc}

\usepackage[utf8]{inputenc}

\usepackage{microtype}

\usepackage{inconsolata}

\usepackage{graphicx}

\title{SW-ASR: A Context-Aware Hybrid ASR Pipeline for Robust Single Word Speech Recognition}

\author{Manali Sharma\thanks{These authors contributed equally to this work.}, Riya Naik\footnotemark[1] 
\and Buvaneshwari G \\
\texttt{manali,riya.n,buvaneshwari@tetranetics.com}\\
Tetranetics Private Limited \\ Mumbai, India}

\begin{document}
\maketitle
\begin{abstract}
    

Automatic Speech Recognition (ASR) for single-word detection is a specialized task that focuses on accurately identifying isolated spoken words. It is especially critical for constrained and open-vocabulary applications in low-resource, communication-sensitive domains such as healthcare and emergency response. Unlike continuous speech ASR, which benefits from surrounding linguistic context, single-word ASR faces challenges due to its minimal context. This is in addition to the issues that are common among continuous speech and single-word ASR such as pronunciation variability, background noise, and speaker diversity. This paper reviews current deep learning approaches and presents a modular framework designed to enhance single-word detection accuracy. The pipeline first applies denoising and volume normalization, then uses a hybrid ASR front end (Whisper + Vosk) with confidence-weighted selection to produce an initial transcription. To address out-of-vocabulary words and low-quality channels, we add a verification layer that can operate in four modes: cosine-embedding similarity, Levenshtein distance, LLM-based matching, and context-guided matching (cosine/LLM with surrounding context). This architecture then integrates with SIP-based telephony stacks, enabling intent-driven functionalities that can be used for cases such as blind call transfers and emergency alerts. For evaluation we combine a public benchmark with platform-specific recordings. We use Google Speech Commands (65k one-second clips across 30 words), and we curate a supplementary dataset that records the same 30 words across real-world channels: cellular voice calls and voice-messaging on WhatsApp, WebChat, and Facebook Messenger. To mirror deployment conditions, we target bandwidth-limited and compressed audio typical of telecommunication networks (e.g., 8 kHz sampling, codec artifacts, noise). Across GSC and the curated platform recordings, the hybrid Whisper + Vosk front end performs best on high-quality audio, while the verification layer yields clear gains on noisier channels. LLM-based matching with contextual prompts consistently reduces word error rate on telephony and WeChat audio, with few-shot prompting providing the strongest improvements among low-quality signals. Context-guided cosine narrows the gap with LLMs in many conditions and can outperform non-contextual methods, offering a favorable accuracy–latency trade-off. Timing analyses show that cosine with context has latency comparable to hybrid/Levenshtein pipelines, while a naïve LLM prompt is costlier; however, adding context + instructions (and few-shot) focuses the LLM enough that its average time approaches cosine in practice. These results suggest that modest verification and context mechanisms can deliver robustness in single-word detection without sacrificing responsiveness required for live telephony actions.
\end{abstract}

\section{Introduction and Related Work}

Automatic Speech Recognition systems have become integral to modern communication platforms, powering voice assistants, customer service automation, emergency response triggers, and command-and-control interfaces \cite{atayero2009implementation}. Although most ASR research efforts focus on continuous speech recognition, an equally important but underexplored area is single-word recognition, where the system must identify and interpret speech input consisting of just one word. Such scenarios are common in voice command interfaces, helpline bots, etc.
The recognition of single words has been studied in the context of keyword identification and isolated word recognition \cite{michaely2017keyword,paul2025isolated}. Traditional approaches rely on mixture models\cite{dines2010measuring,nainan2016comparison} constrained by domain-specific vocabularies and lacking the adaptability required for open vocabulary. These systems depend on cloud-based inference with massive compute availability, limiting their applicability in resource-constrained or latency-sensitive environments.

With the advent of deep learning, neural network techniques have improved recognition performance. CNNs\cite{pan2020asapp} have been applied to spectrogram-based word classification, while LSTM and GRUs have been used to model temporal dependencies and enhance robustness \cite{passricha2019hybrid,tonk2024automatic}. Transformer architectures using Wav2Vec and Whisper have shown state-of-the-art performance by learning contextual representations \cite{radford2023robust,baevski2020wav2vec}. While these have addressed many challenges, such as pronunciation variation, background noise, and speaker diversity, single-word signals still face a unique limitation: the absence of surrounding context. Our single word (SW-ASR) operates in isolation, without grammatical or semantic cues to support inference, making accurate detection difficult and even worse when dealing with low-quality audio signals.

In this paper, we propose a modular framework for SW-ASR for both constrained- and open-vocabulary scenarios. Our architecture integrates deep learning-based acoustic models with SIP-compatible telephony systems, enabling accurate recognition and immediate response actions, such as blind call transfers, emergency alerts, and role-based routing. We evaluate our system on real-world platforms, including cellular networks and voice messaging applications such as WhatsApp\footnote{\url{https://www.whatsapp.com/}}, WeChat\footnote{\url{https://www.wechat.com/}}, and Facebook Messenger\footnote{\url{https://www.facebook.com/messenger/}}, and demonstrate substantial improvements in recognition accuracy, even under noisy and resource-limited conditions.

\textbf{Case Study: Real-World Applications of Single-Word ASR}
The SW-ASR framework proposed in this paper is developed to address the challenges observed in various public-facing deployments, such as \\
(1) A citizen-facing chatbot was deployed on WhatsApp and social media platforms to report power outages in a region with more than 130 million people and a literacy rate of only 63\%. Many users, often unable to type in their native script, chose to record voice messages with short phrases like '\textit{bijli gayi}' ('electricity gone'). Initial deployments showed that the system underperformed in detecting such informal speech without adaptation to multilingual, noisy, and informal inputs, requirements that directly shaped the improvements discussed in this document.\\
(2) An emergency helpline received over 130,000 calls, with fewer than 0.03\% being genuine. Most were pocket dials, pranks, or non-emergencies. To manage this, an ASR-based filtration system was deployed to detect urgent keywords like "\textit{help}," "\textit{fire}," or "\textit{ambulance}." It had to operate reliably in noisy, high-stress conditions to ensure genuine emergencies weren’t missed.\\
(3) A secure telephony environment in which personnel could initiate call transfers simply by speaking the name of a colleague. The ASR system needs to accurately recognize a finite but dialect-sensitive set of proper nouns, enabling spoken inputs to replace memorized extensions in complex, distributed environments. Although we quantitatively evaluated the model using publicly available datasets, the system architecture and training choices were directly informed by the demands and failure points from these real-world applications.

\section{Experimental Setup}
\label{sec:expt}


\subsection{Data}
In our experimental investigation, we focus on the publically available \textbf{Google Speech Commands}\footnote{\url{https://research.google/blog/launching-the-speech-commands-dataset/}} (GSC) dataset \cite{sainath2015convolutional}. The GSC dataset consists of approximately 65,000 one-second audio recordings that cover 30 distinct words. To evaluate generalization across platforms, we recorded the same 30 words on multiple channels, each paired with its textual label for training and evaluation.

\subsubsection{Data Curation}
GSC consists of exceptionally high-quality audio recordings characterized by articulate speech and minimal periods of silence or distortion, a level of quality that is uncommon in real-world datasets. In contrast, speech data in real-world applications often originates from bandwidth-limited environments, such as telecommunication networks, where audio is sampled at 8 kHz and subject to compression and noise.
To reflect this variability, we curated a supplementary dataset by recording the same 30 words via WhatsApp, cellular calls, WeChat, and Facebook voice messaging. Our annotation team of native speakers meticulously crafted speech samples for every word on these diverse platforms to encapsulate nuanced differences in audio quality. We gathered ten audio recordings per word from each participant, accumulating a total of 300 audio recordings per platform, which collectively amounted to 1,200 speech samples spanning varied real-world conditions.


\subsection{System Model and Design}
\begin{figure}
    \centering
    \includegraphics[width=0.47\textwidth]{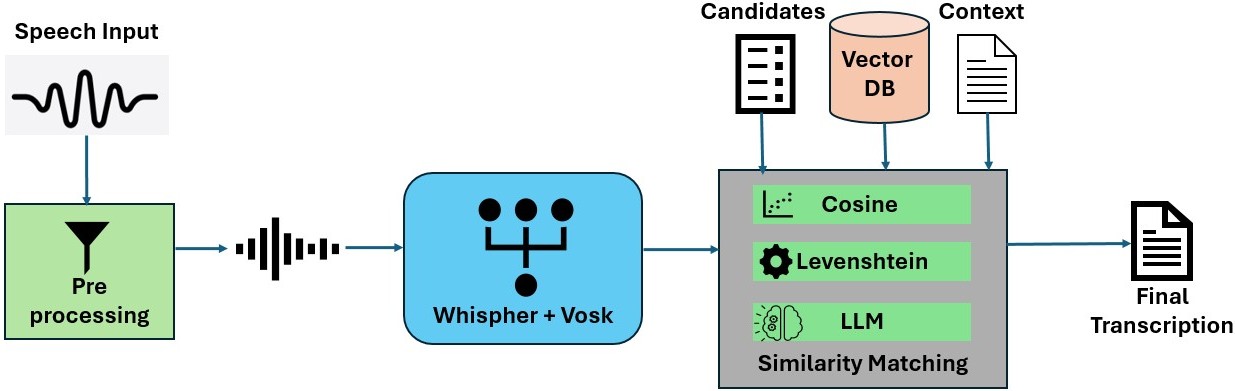}
    \caption{SW-ASR: (1) The single-word speech audio processed to improve quality. (2)Hybrid: Whisper \& Vosk used to generate initial transcript. (3) Candidate matching performed using three approaches. (4) Contextual Concatenation for Generalization.}
    \label{fig:pip}
\end{figure}

The system framework is designed with key components depicted in Figure \ref{fig:pip}. The raw audio input undergoes preprocessing consisting noise reduction and volume normalization. A hybrid combination of OpenAI's Whisper\footnote{\url{https://huggingface.co/openai/whisper-base}} and Vosk\footnote{\url{https://alphacephei.com/vosk/models}} is used for ASR. The order of processing through this hybrid model begins with Whisper for accuracy followed by Vosk for improved single-word detection via its phoneme-trained backbone. To further improve recognition context is injected into each matching alternative. The working and formulation of each key component is detailed below:


\subsection{Initial Audio Transcription}
The audio input \( X \) is first pre-processed and transcribed using a hybrid approach that integrates the outputs of Whisper and Vosk, weighted by their respective confidence scores.  
\(T_W, C_W = f_W(X)\) and \(T_V, C_V = f_V(X)\)
\noindent
where \( T_W \) is the Whisper transcription output, and \( f_W \) represents the Whisper model and \( T_V \) is Vosk’s transcription and \( f_V \) represents the Vosk model. \( T_{\text{M}} =  T_W \text{if } C_W \geq C_V \text{ and } C_W \geq \tau \text{ otherwise } T_V \). Here \(\tau\) is a predefined confidence threshold. Our selection strategy ensures that the transcription with the highest confidence across the ensemble is chosen.

\subsection{Similarity Matching}
Relying solely on the $T_M$ as the final output can be problematic in scenarios where the word in question is unique, falls outside the training vocabulary, or is derived from low-quality audio signals. To address this and improve the overall robustness of the system, we incorporate a post-processing step for known vocabulary cases. This involves comparing the initial transcription against a predefined set of target words using a suitable similarity metric. This approach helps refine the transcription output by aligning it accurately with the expected vocabulary.

\subsubsection{Cosine Similarity}
\label{sec:cosine}
In the first approach, we convert both the initial transcription and target words into vectors using a pre-trained embedding model and match them using cosine similarity\cite{laskar2020contextualized}. These embeddings capture semantic and syntactic relationships between words. We compute cosine similarity between the vector of the initial transcription and each target word vector, and map the initial transcription to the target word with the highest similarity.

\subsubsection{Levenshtein Distance}
\label{sec:lev}
Cosine similarity relies on semantic embeddings of a pre-trained model and may not work reliably on out-of-vocabulary or domain-specific words. For this reason, we test the similarity matching using Levenshtein distance \cite{zhang2017research}. It operates on a character level, directly measuring how many edits (insertions, deletions, substitutions) are needed to turn one word into another. This feature makes it efficient for single-word matching without using any embedding model, handling orthographically different words.

\subsubsection{LLM Matching}
Cosine and Levenshtein are effective to some extent, but they do not account for phonetic similarity. They can fail when two words sound alike but differ in terms of spelling and semantics, such as  "\textit{phone}" and "\textit{fone}". To improve on this for the known vocabulary cases, we prompt LLM with initial transcription and a list of target words to select the most plausible match based on semantic understanding and phonetics. This enables intelligent corrections, especially in cases where the transcription is ambiguous or incomplete.

\subsubsection{Context Guided Matching}
This approach combines cosine similarity and LLMs with contextual information to improve accuracy further and reduce the word error rate. Rather than processing words in isolation, embeddings are calculated with surrounding context for better semantic relevance. Although cosine with context shows substantial improvement, it can sometimes fail in cases where there can be two alternatives. For example, if the context is “\textit{I have two pets, a dog and a cat}” and the target word is “\textit{cat}”, cosine similarity may incorrectly match it to “\textit{dog}”. In such cases, LLMs perform better by using contextual prompts to understand broader discourse, enabling more accurate selection based on meaning, grammar, and coherence. This hybrid approach with similarity matching ensures a balanced trade-off between accuracy and computational efficiency.




\section{Results and Discussion}
We now discuss the evaluation and comparison of pipelines drawn in section \ref{sec:expt} for the GSC dataset alongside platform-specific augmentations. We label each pipeline as: \textbf{Hybrid} for initial hybrid transcription using Whisper \& Vosk, \textbf{CS} for cosine similarity, \textbf{LS} for Levenshtein similarity, \textbf{LLM} for LLM-based similarity, \textbf{CS + C} for context-embedded cosine similarity, \textbf{LLM + C} for context-embedded LLM similarity, and \textbf{LLM + C + FS} for context-embedded LLM similarity with few-shot prompting.

We assess the single word recognition capabilities of each pipeline across datasets using accuracy, word error rate (WER), and time taken. As tabulated in Tables \ref{tab:acc} and \ref{tab:wer}, the initial hybrid method performs better on the GSC dataset due to its high quality as compared to other platforms with low-quality signals. We observe the advantages of using similarity matching in such scenarios, specifically on Facebook data. Using standard similarity measures such as \ref{sec:cosine} and \ref{sec:lev} is computationally efficient, but these methods have limitations: Although the Levenshtein distance is similar to cosine matching, it has the potential to deviate as the context increases. Cosine distance takes semantic similarity, but it does not inherently take phonetics into account. Thus, if the initial transcription is semantically distinct but phonetically overlapping, it can be mismatched. 
\noindent
LLMs are trained on a huge amount of data, which can be leveraged for similarity matching. We observed Llama-4-Scout\footnote{\url{https://ai.meta.com/blog/llama-4-multimodal-intelligence/}} here to measure the matching performance. We first test it with a naive prompt that instructs the LLM to map the transcription to the word in the target set. LLM matching shows a significant improvement over standard measures; even in the telephony setting, the WER is substantially reduced. Unlike standard measures, LLMs take into account various features of a word (syntactic, semantic, and phonetic) to perform the mapping. 
\begin{table}[h!]
\centering
\tiny
\begin{tabular}{|l|c|c|c|c|c|c|c|}
\hline
\textbf{Dataset} & \textbf{Hybrid} & \textbf{CS} & \textbf{LS} & \textbf{LLM} & \textbf{CS } & \textbf{LLM } & \textbf{LLM } \\
&  &  &  &  & \textbf{+ C} & \textbf{+ C} & \textbf{+ C + FS} \\
\hline
GSC & 0.49 & 0.59 & 0.61 & 0.66 & 0.65 & 0.77 & 0.86 \\
GSC-Wh   & 0.28 & 0.38 & 0.31 & 0.42 & 0.47 & 0.68 & 0.69 \\
GSC-T & 0.11 & 0.26 & 0.23 & 0.35 & 0.36 & 0.51 & 0.51 \\
GSC-We & 0.20 & 0.34 & 0.43 & 0.53 & 0.49 & 0.67 & 0.73 \\
GSC-FB & 0.42 & 0.44 & 0.58 & 0.61 & 0.47 & 0.70 & 0.71 \\
\hline
\end{tabular}
\caption{Accuracy Comparison across datasets}
\label{tab:acc}
\end{table}

\begin{table}[h!]
\centering
\tiny
\begin{tabular}{|l|c|c|c|c|c|c|c|}
\hline
\textbf{Dataset} & \textbf{Hybrid} & \textbf{CS} & \textbf{LS} & \textbf{LLM} & \textbf{CS } & \textbf{LLM } & \textbf{LLM } \\
&  &  &  &  & \textbf{+ C} & \textbf{+ C} & \textbf{+ C + FS} \\
\hline
GSC & 0.59 & 0.41 & 0.39 & 0.35 & 0.35 & 0.24 & 0.15 \\
GSC-Wh & 0.96 & 0.65 & 0.70 & 0.58 & 0.53 & 0.33 & 0.32 \\
GSC-T & 1.00 & 0.77 & 0.80 & 0.65 & 0.64 & 0.49 & 0.48 \\
GSC-We & 1.00 & 0.67 & 0.57 & 0.48 & 0.52 & 0.33 & 0.27 \\
GSC-FB & 0.81 & 0.57 & 0.44 & 0.41 & 0.53 & 0.30 & 0.29 \\
\hline
\end{tabular}
\caption{WER Comparison across datasets}
\label{tab:wer}
\end{table}

\textbf{Context}. The ASR models are trained on long-context sentences and thus fail to recognize isolated single-word queries. But these SW queries are often instructions/commands to perform an action or an answer or choice to a question. This observation makes it suitable for SW transcriptions to be concatenated with relevant context to boost the performance. We observe that having additional context for matching enhances the performance. CS + C gives comparable results to using LLM with context. And marginally better on WhatsApp and Telephony signals. We also see that LLM with a contextual prompt with additional instructions understands the context and maps the target word more efficiently with at least 10\% in WER and, achieves the best results for low-quality telephony and WeChat signals. This approach can be further improved using state-of-the-art prompting strategies. For our experiment, we adopt few-shot prompting. It is evident from the results that few-shot helps LLMs to understand the data and handle marginal cases, specifically when multiple single-word answers/commands are possible. 

\begin{figure}[h]
    \centering
    \includegraphics[width=0.40\textwidth]{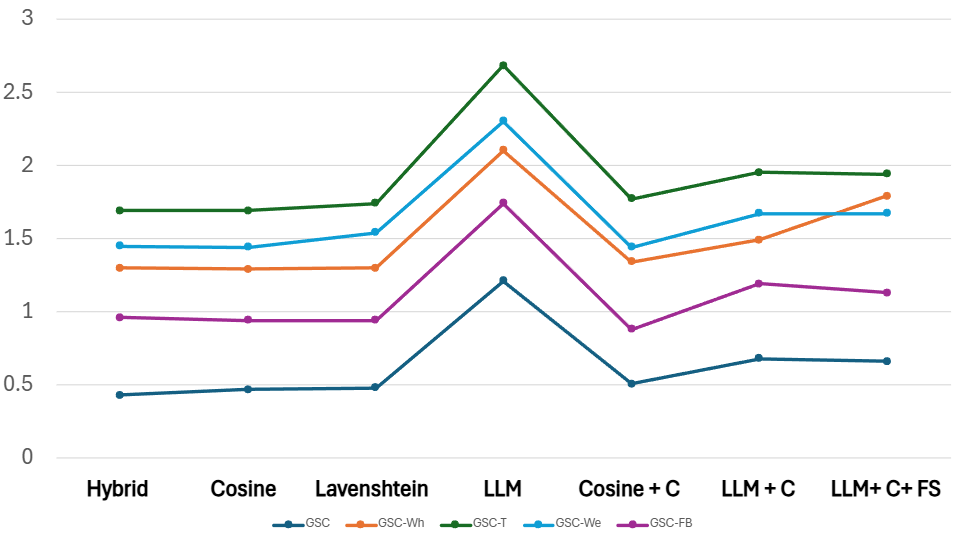}
    \caption{Average time taken across datasets for each approach}
    \label{fig:time}
\end{figure}

\noindent
It's always a tradeoff to prioritize between performance and latency when we incorporate transformers and LLMs. Figure \ref{fig:time} shows the average time taken by each pipeline alternative, and we can witness that cosine similarity with context amounts to similar time as compared to hybrid and LS. LLM with naive prompt takes the highest, this is due to the nature of LLM to consider a broad range of features and generalization. When clubbed with better prompt and context, LLM interestingly consumes a similar amount as CS. Since, based on instructions, context, and a few shots, it can narrow its focus and perform mapping easily.

\section{Conclusion}
In this paper, we present a modular framework for Single-Word Automatic Speech Recognition (SW-ASR) applicable to both constrained and open-vocabulary settings. We explore multiple strategies to enhance the accuracy and robustness of SW-ASR systems. To ensure generalizability and practical relevance, we evaluate our methods on audio data across diverse platforms. Our hybrid approach, incorporating similarity-based matching, demonstrates consistent performance improvements across these environments. The findings highlight the feasibility of deploying SW-ASR systems in real-world scenarios, particularly where computational or data limitations preclude extensive fine-tuning or retraining of acoustic models. 

\section*{Limitations}
Our single-word (SW-ASR) framework was tested on WhatsApp, WeChat, Facebook Messenger, cellular calls, and recorded audio, covering both synchronous and asynchronous voice interactions. However, several constraints limit its generalizability across platforms and contexts.
First, platform-specific audio characteristics (e.g., compression rates, codecs, noise artifacts) introduce variability. WhatsApp voice notes may suffer from compression distortions, WeChat uses different sampling rates, and phone calls experience telecom-induced bandwidth constraints and channel noise. Each platform required targeted model adaptation.
Second, dialect and language diversity significantly affect performance. While the system supports multilingual and dialect-sensitive inputs, generalization to low-resource or code-switched languages may require domain-specific tuning and additional data.
Third, speaker conditions affect generalizability. The framework performs best on single-speaker utterances, with reduced accuracy on multi-speaker or conversational inputs.
Finally, despite strong evaluation on public datasets, edge cases and atypical conditions (e.g., extreme background noise, overlapped speech, novel dialects) remain underrepresented. This work can be enhanced for language coverage, introduce lightweight on-device adaptation, and fine-tune for cross-platform deployment.

\section*{Ethics Statement}

Our proposed research adheres to the ACL Code of Ethics and involves no risk to individuals or communities. All supplementary speech data was generated in-house. The primary data collector and annotator is one of the authors. Recordings were limited to predefined, non-identifying single-word utterances. Annotators were fairly compensated. No personal or sensitive information was collected. Given the intended use in public service domains like emergency response, we acknowledge dual-use concerns and have designed the system with human fallback mechanisms to prevent over-reliance on automation. No trained models are released publicly. By curating multilingual, platform-specific data, this work aims to advance fairness and generalizability in speech recognition for low-resource and real-world deployment settings.

\bibliography{custom}

\end{document}